\pdfoutput=1

\RequirePackage{fix-cm}

\documentclass[twocolumn]{svjour3}

\smartqed

\usepackage{graphicx}
\usepackage{amssymb,amsmath}
\usepackage{txfonts}
\usepackage{natbib}
\usepackage{hyperref}

\hypersetup{pdftitle = The title of my PDF, pdfauthor = My name, pdfsubject= The subject, pdfkeywords = keyword1 keyword2 keyword3}
\hypersetup{colorlinks = true, linkcolor = blue, anchorcolor = red, citecolor = blue, filecolor = red, pagecolor = red, urlcolor = blue}

\journalname{Astrophysics and Space Science}

\begin{document}

\title{Basins of attraction of equilibrium points in the planar circular restricted five-body problem}

\author{Euaggelos E. Zotos \and Md Sanam Suraj}

\institute{Euaggelos E. Zotos: \at
             Department of Physics, School of Science, \\
             Aristotle University of Thessaloniki, \\
             GR-541 24, Thessaloniki, \\
             Greece \\
             \email{evzotos@physics.auth.gr}
         \and
           Md Sanam Suraj: \at
             Department of Mathematics, SGTB Khalsa College, \\
             University of Delhi, North Campus, \\
             New Delhi-110007, \\
             India \\
             \email{mdsanamsuraj@gmail.com}
}

\date{Received: 7 October 2017 / Accepted: 21 December 2017 / Published online: 02 January 2018}

\titlerunning{Basins of attraction of equilibrium points in the planar circular restricted five-body problem}

\authorrunning{E.E. Zotos \& Md S. Suraj}

\maketitle

\begin{abstract}

We numerically explore the Newton-Raphson basins of convergence, related to the libration points (which act as attractors), in the planar circular restricted five-body problem (CR5BP). The evolution of the position and the linear stability of the equilibrium points is determined, as a function of the value of the mass parameter. The attracting regions, on several types of two dimensional planes, are revealed by using the multivariate version of the classical Newton-Raphson iterative method. We perform a systematic investigation in an attempt to understand how the mass parameter affects the geometry as well as the degree of fractality of the basins of attraction. The regions of convergence are also related with the required number of iterations and also with the corresponding probability distributions.


\end{abstract}

\section{Introduction}
\label{intro}

In celestial mechanics, dynamical systems of $N$-body problem have been comprehensively inquired for a long period of time. Especially, the simplified models of few-body problem, such as the circular restricted three-body problem \citep{S67} and the circular restricted four-body problem, are used to determine the dynamical properties found in the Solar System. These problems have plenty of applications in scientific research, not only in planetary physics, Solar System and galactic dynamics but also in various fields of astrodynamics and astrophysics as well. Also, several modifications are proposed by introducing perturbing terms to the effective potential to understand the motion of the infinitesimal mass in the problem of three and four bodies.

The restricted few-body problem is, beyond any doubt, the best tool for modeling the motion of test particles in several types of planetary systems in our Solar System. In particular, knowing the equilibrium points (also known as Lagrange points) of a system is an issue of paramount importance since they are very important especially for plotting the trajectories of spacecrafts. For example, a large number of NASA missions have been, or will be, sent to saddle point $L_1$ because this position is ideal for making observations of the Sun-Earth system, since there the gravitational attraction of the Earth partially cancels the gravitational attraction of the Sun. Moreover the $L_2$ saddle point can be thought as a gateway for bodies (e.g., dangerous asteroids that might collide with the Earth) entering the region in the Earth-Moon system. In addition, as it is well known, the Trojan asteroids can be found near the triangular points $L_4$ and $L_5$. On this basis, knowing the equilibrium points of a system, and of course the corresponding basins of convergence, give us very important information regarding the most intrinsic properties of the dynamical system.

A large number of research works is devoted on the existence of equilibrium points \citep[e.g.,][]{KC86}, their stability \citep[e.g.,][]{A12}, the periodic orbits around the equilibrium points \citep[e.g.,][]{AES12} and the basins of attraction, associated with the equilibrium points in the restricted problem of three bodies \citep[e.g.,][]{Z16a}. The effects of a radiating primary body \citep[e.g.,][]{KP78}, oblateness of the primaries \citep[e.g.,][]{SSR75} as well as small perturbations in Coriolis and centrifugal forces \citep[e.g.,][]{BH78} have been included in the same problem. The same aspects have been also investigated in the case of the four-body problem \citep[e.g.,][]{AUHS15a,AUHS15b,AUHS16,AUHS17,KAE06,MASB16,P16,PP13,SAA17,SH14,Z16b,Z17a}.

A number of articles have been published on planar central configurations of $N$-bodies mainly for $N = 4, 5$ and 7. The study of the motion of $N$ point masses moving under their mutual gravitational attraction governed by Newton's gravitational law is always referred as classical $N$-body problem. The central configurations are authoritative in the $N$-body problem since they allow us to obtain the homographic solutions, such as the configuration of the $N$-bodies at any particular point of time, with respect to the inertial barycentric system, which stay similar to itself as the time changes. It is well known that the first three collinear homographic solutions and two another homographic solutions for $N = 3$, known as the equilateral triangle solutions, were found by Euler in 1767 and by Lagrange in 1772, respectively.

A series of research articles are available on the central configuration for $N > 3$ bodies, and the analysis of these configuration is still an interesting and attracting field of research \citep[e.g.,][]{ARL13,EC16,H05,HS07,LM09,M10,MF13,SA13}.

It was \citet{O88} who introduced the gravitational five-body problem to discuss the motion of the fifth body of negligible mass, in comparison to other bodies. He has supposed that the three bodies with equal masses revolve on the same plane, around their gravitational center in circular orbit under their mutual gravitational pull. In addition, a mass of $\beta > 0$ times the mass of one of the three primary bodies is situated at the centre of mass. For $\beta = 0$, this particular case of the restricted five-body problem is reduced to the restricted four-body problem. He has obtained nine libration points in total, where three of them become stable when $\beta > 43.18$, while for smaller values all the libration points are linearly unstable.

\citet{KMV11} discussed the restricted rhomboidal five-body problem and showed that the total number of equilibrium points depends on the ratio of the semi-diagonals and there can be eleven, thirteen or even fifteen libration points, which are all unstable. In the same vein, \citet{MV13} studied the spatial restricted rhomboidal five-body problem and also studied the horizontal stability of its periodic solutions. In this coplanar problem of five bodies, the four primaries move two by two in circular orbits, while the center of mass is taken as the origin, so that the primaries always maintain a rhombus configuration.

\citet{PK07} numerically explored the restricted five-body problem, when some or all of the primary bodies are sources of radiation, thus extending the work of Oll\"{o}ngren. They revealed that the number of collinear libration points of the system depends on the radiation factors and of course on the mass parameter. Recently, \citet{GYS17} studied the motion of the fifth body with infinitesimal mass in the axisymmetric restricted five-body problem, when the four primaries are maintaining the axisymmetric central configurations discussed in \citet{EC16}. They revealed that there exist at most fifteen equilibrium points for convex configuration, while thirteen libration points exist for concave configuration, depending upon the angle coordinates. Moreover, they showed that all the libration points are linearly unstable for all the combinations of the angle coordinates, in three axisymmetric configurations.

In the present work we shall use the mathematical model of the restricted five-body problem introduced in \cite{O88}. Our aim is to reveal how the geometry as well as the shape of the Newton-Raphson basins of attraction are influenced by the mass parameter. Our paper has the following structure: the most important properties of the dynamical system are presented in Section \ref{mod}. The parametric evolution of the position as well as of the stability of the equilibrium points is investigated in Section \ref{param}. The following Section contains the main numerical results, regarding the evolution of the Newton-Raphson basins of convergence. In Section \ref{bee} we demonstrate how the basin entropy evolves, as a function of the mass parameter, while in the next section we provide the main concluding remarks. Our paper ends with Section \ref{fut}, where we emphasize related aspects which will be incorporated in future works.

\section{Description of the mathematical model}
\label{mod}

In the circular restricted five-body problem (CR5BP) the four primaries, $P_0$, $P_1$, $P_2$, and $P_3$, move on coplanar circular orbits around their common center of gravity. We assume that the fifth body has a significantly smaller mass related to the masses of the primary bodies ($m^{*} \ll m_0, m_1, m_2, m_3$). On this basis, the fifth body acts as an infinitesimal test particle and therefore it does not influence the circular motion of the four primaries.

\begin{figure}[!t]
\centering
\resizebox{\hsize}{!}{\includegraphics{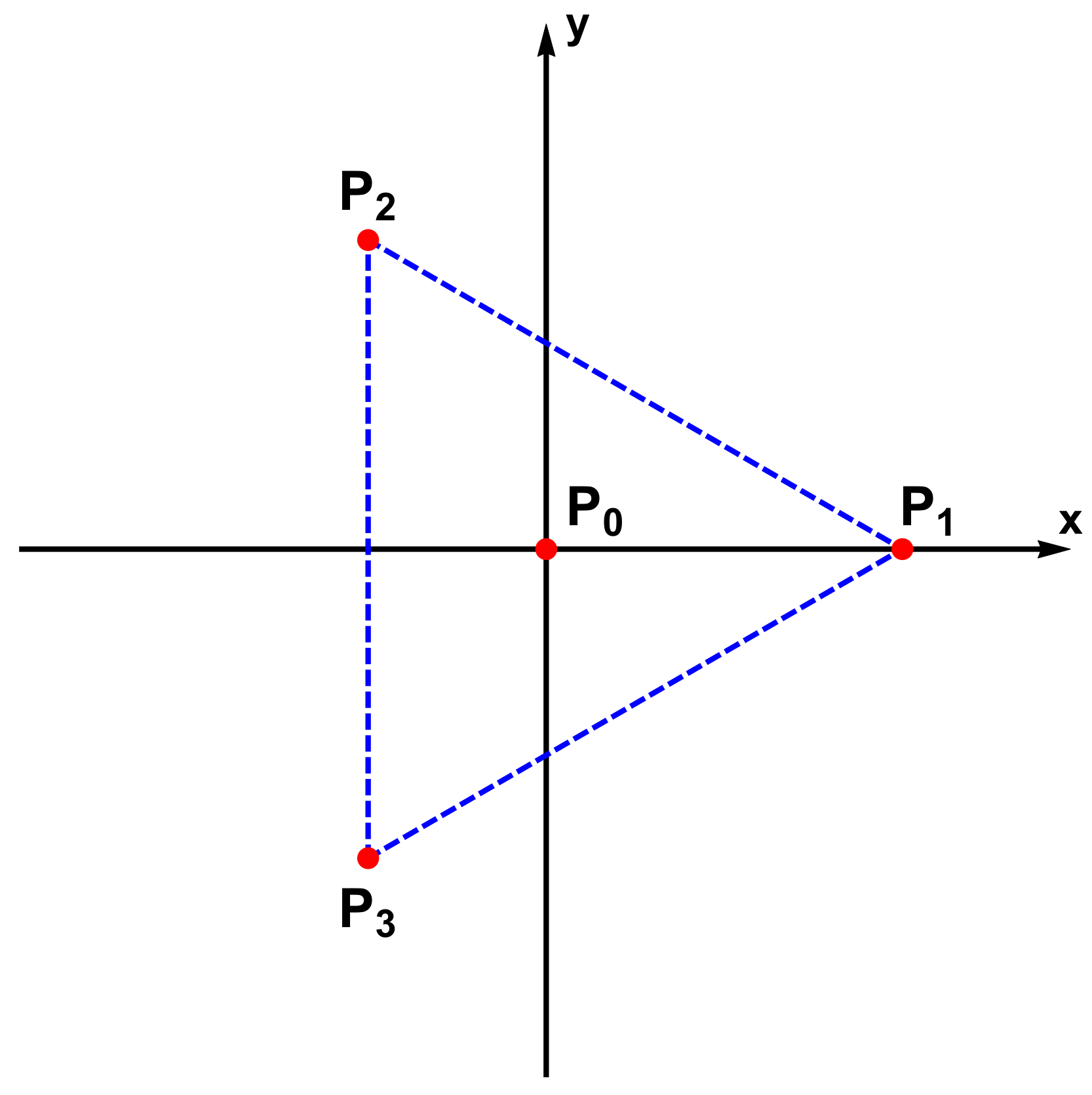}}
\caption{The planar configuration of the circular restricted five-body problem (CR5BP). The positions of the four primary bodies are indicated by red dots. The three bodies with equal masses $m$ are located at the vertices of an equilateral triangle (dashed blue lines), while the fourth primary, with mass $\beta~m$, is located at the center of the equilateral triangle. (Color figure online).}
\label{conf}
\end{figure}

For modeling the planar motion of the fifth body we choose a rotating frame of reference in which the origin $O(0,0)$ coincides with the center of the mass of the primary bodies. The dimensionless masses of the primaries are $m_0 = \beta~m$, $m_1 = m_2 = m_3 = m = 1$, while the positions of their centers are: $(x_0,y_0) = (0,0)$, $(x_1,y_1) = (1/\sqrt{3},0)$, $(x_2,y_2) = (-x_1/2,1/2)$, and $(x_3,y_3) = (x_2,-y_2)$. In Fig. \ref{conf} we present the planar configuration of the CR5BP. It is seen that the three bodies of mass $m$ are located at the vertices of an equilateral triangle with side $a = 1$, while the fourth primary, with mass $\beta~m$, is located at the center of the equilateral triangle.

Looking at Fig. \ref{conf} it becomes evident that the CR5BP admits a $2\pi/3$ symmetry. Indeed, if we rotate the primary symmetry $x$ axis $(y = 0)$, successively through an angle of $2\pi/3$ we obtain the two additional lines of symmetry of the system, that is $y = \sqrt{3}$ and $y = -\sqrt{3}$.

According to \citet{O88} the time-independent effective potential function of the CR5BP is
\begin{equation}
\Omega(x,y) = k \sum\limits_{i=0}^3 \frac{m_i}{r_i} + \frac{1}{2}\left(x^2 + y^2 \right),
\label{pot}
\end{equation}
where
\begin{equation}
k = \frac{1}{3\left(1 + \beta~\sqrt{3}\right)},
\label{par}
\end{equation}
while
\begin{equation}
r_i = \sqrt{\left(x - x_i\right)^2 + \left(y - y_i\right)^2}, \ \ \ i = 0,1,2,3,
\label{dist}
\end{equation}
are the distances of the fifth body from the four primaries. It is interesting to note that when $\beta = 0$ (which means that the central primary $P_0$ is absent) the system is automatically reduced to the circular equilateral restricted four-body problem.

The equations of motion, in synodic coordinates, of the test particle read
\begin{align}
\ddot{x} - 2 \dot{y} &= \frac{\partial \Omega}{\partial x}, \nonumber\\
\ddot{y} + 2 \dot{x} &= \frac{\partial \Omega}{\partial y},
\label{eqmot}
\end{align}
where
\begin{align}
\Omega_x(x,y) &= \frac{\partial \Omega}{\partial x} = - k \sum\limits_{i=0}^3 \frac{m_i\left(x - x_i\right)}{r_i^3} + x, \nonumber\\
\Omega_y(x,y) &= \frac{\partial \Omega}{\partial y} = - k \sum\limits_{i=0}^3 \frac{m_i\left(y - y_i\right)}{r_i^3} + y.
\label{der1}
\end{align}
In the same vein, the partial derivatives of the second order, which will be needed later for the multivariate Newton-Raphson iterative scheme, are the following
\begin{align}
\Omega_{xx}(x,y) &= \frac{\partial^2 \Omega}{\partial x^2} = k \sum\limits_{i=0}^3 \frac{m_i \left(3 \left(x - x_i\right)^2 - r_i^2\right)}{r_i^5} + 1, \nonumber\\
\Omega_{xy}(x,y) &= \frac{\partial^2 \Omega}{\partial x \partial y} = 3 k \sum\limits_{i=0}^3 \frac{m_i \left(x - x_i\right)\left(y - y_i\right)}{r_i^5}, \nonumber\\
\Omega_{yx}(x,y) &= \frac{\partial^2 \Omega}{\partial y \partial x} = \Omega_{xy}(x,y), \nonumber\\
\Omega_{yy}(x,y) &= \frac{\partial^2 \Omega}{\partial y^2} = k \sum\limits_{i=0}^3 \frac{m_i \left(3 \left(y - y_i\right)^2 - r_i^2\right)}{r_i^5} + 1.
\label{der2}
\end{align}

The Jacobi integral of motion, corresponding to the system of differential equations (\ref{eqmot}) is given by
\begin{equation}
J(x,y,\dot{x},\dot{y}) = 2\Omega(x,y) - \left(\dot{x}^2 + \dot{y}^2 \right) = C,
\label{ham}
\end{equation}
where $\dot{x}$ and $\dot{y}$ are the velocities, corresponding to coordinates $x$ and $y$, respectively, while $C$ is the numerical value of the Hamiltonian which is conserved.

\section{Parametric evolution of the equilibrium points}
\label{param}

For comparison reasons we can easily define a mass parameter $\mu = 1/(1 + \beta)$, similar to the mass parameter of the classical restricted three-body problem. Therefore we have that $\mu (0, 1]$ when $\beta \in [0, \infty)$.

The necessary and sufficient conditions, which must be fulfilled for the existence of coplanar equilibrium points, are
\begin{equation}
\dot{x} = \dot{y} = \ddot{x} = \ddot{y} = 0.
\label{lps0}
\end{equation}
The corresponding coordinates $(x,y)$ of the libration points can be determined by solving numerically the system of the first order derivatives
\begin{equation}
\Omega_x(x,y) = 0, \ \ \ \Omega_y(x,y) = 0.
\label{lps}
\end{equation}

\begin{figure*}[!t]
\centering
\resizebox{\hsize}{!}{\includegraphics{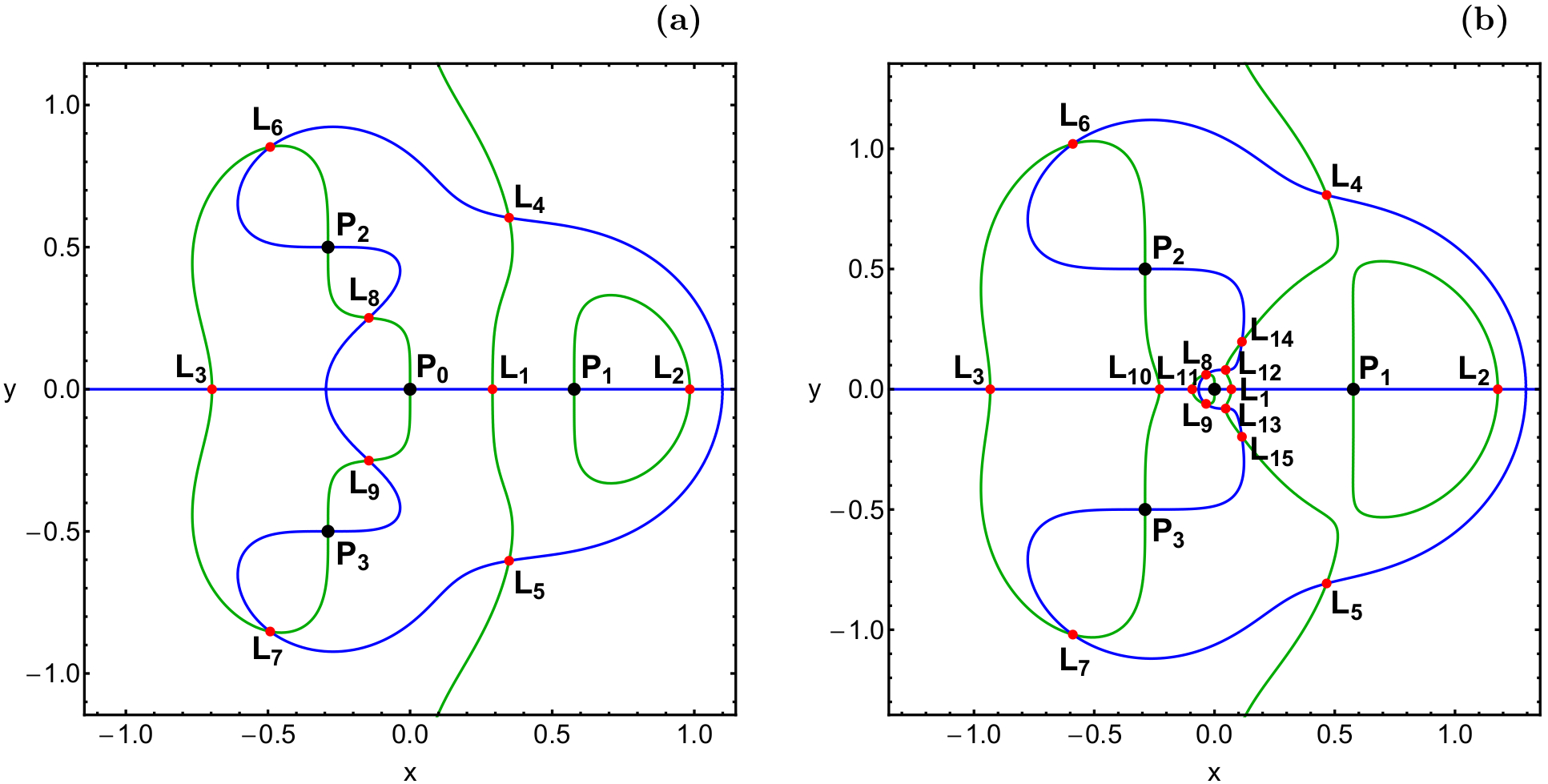}}
\caption{Positions (red dots) and numbering of the equilibrium points $(L_i, \ i=1,...,15)$ through the intersections of $\Omega_x = 0$ (green) and $\Omega_y = 0$ (blue), when (a-left): $\mu = 0.5$ (nine equilibrium points), and (b-right): $\mu = 0.995$ (fifteen equilibrium points). The black dots denote the centers $(P_i, \ i=0,1,2,3)$ of the primaries. (Color figure online).}
\label{lgs}
\end{figure*}

The total number of libration points in the CR5BP is a function of the mass parameter $\mu$. In particular
\begin{itemize}
  \item When $\mu \in (0, 0.98617275]$ there exist nine equilibrium points: three collinear and six non-collinear points (see panel (a) of Fig. \ref{lgs}).
  \item When $\mu \in [0.98617276, 1)$ there exist fifteen equilibrium points: five collinear and ten non-collinear points (see panel (b) of Fig. \ref{lgs}).
  \item When $\mu = 1$ there are ten libration points, as in the restricted equilateral four-body problem with three equal masses.
\end{itemize}
The value $\mu^{*} = 0.98617276$ is a critical value of the mass parameter, since it delimits the point where the number of the equilibrium points changes.

The intersections of the nonlinear equations $\Omega_x = 0$, and $\Omega_y = 0$ define the positions of the equilibrium points. Fig. \ref{lgs}(a-b) illustrates how these equations pinpoint, in each case, the location of the libration points, when (a): $\mu = 0.5$ and (b): $\mu = 0.995$. In the same diagram we explain the numbering, $L_i, \ i=1,...,15$, of all the equilibrium points.

\begin{figure}[!t]
\centering
\resizebox{\hsize}{!}{\includegraphics{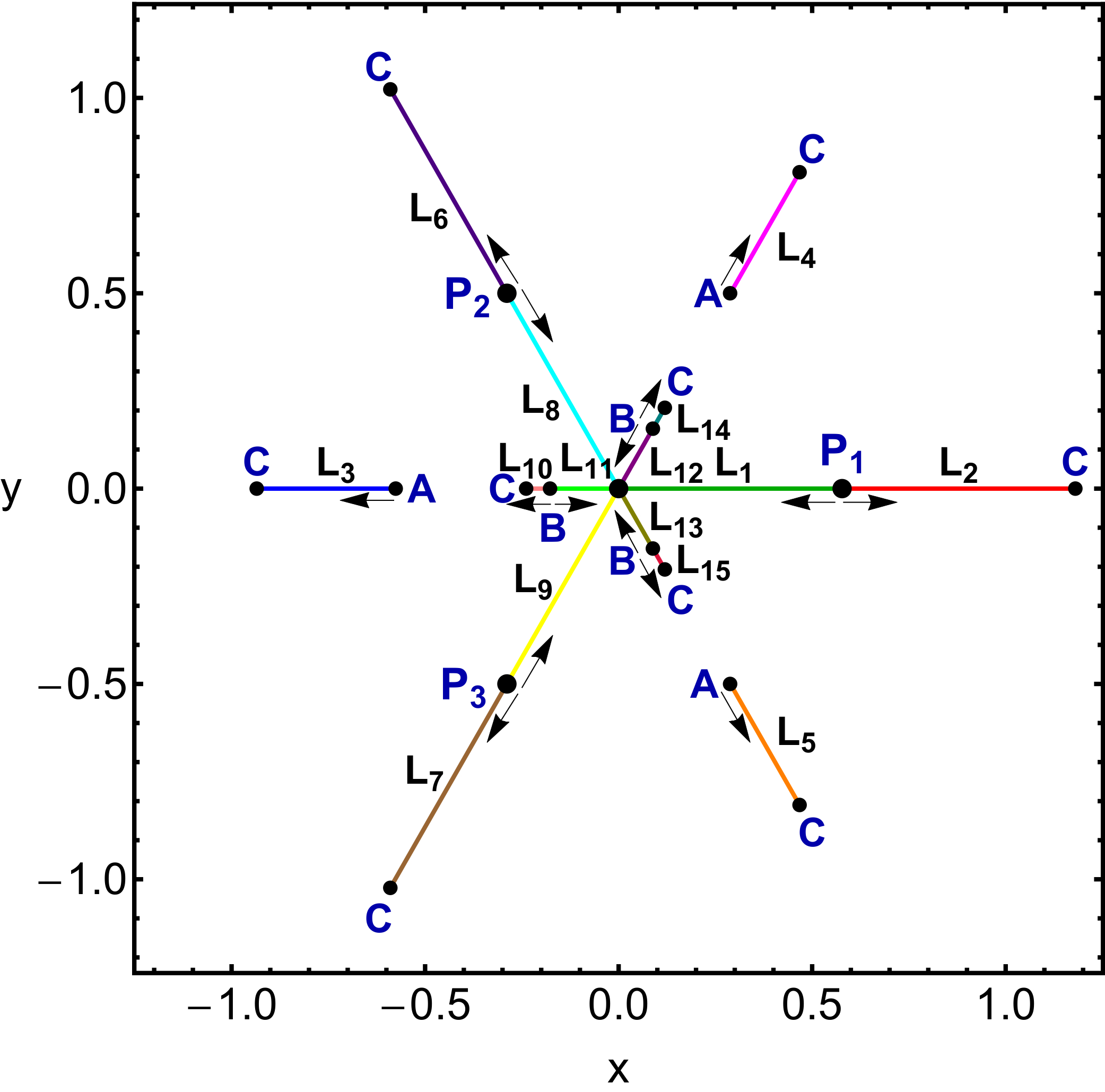}}
\caption{The parametric evolution of the positions of the equilibrium points, $L_i$, $i=1,...,15$, in the CR5BP, when $\mu \in (0, 1]$. The arrows indicate the movement direction of the equilibrium points as the value of the mass parameter increases. The big black dots pinpoint the fixed centers of the primaries, while the small black dots (points A, B, and C) correspond to $\mu \to 0$, $\mu = \mu^{*}$, and $\mu = 1$, respectively. (Color figure online).}
\label{evol}
\end{figure}

In Fig. \ref{evol} we present the parametric evolution of the positions of the equilibrium points, on the configuration $(x,y)$ plane, when $\mu \in (0,1]$. It is seen that as soon as $\mu$ is just above zero three pairs of equilibrium points emerge from the centers $P_1$ ($L_1$ and $L_2$), $P_2$ ($L_6$ and $L_8$), and $P_3$ ($L_7$ and $L_9$) of the primaries. When $\mu = \mu^{*}$ three additional pairs of libration points appear. As the value of mass parameter grows the equilibrium points evolve following two different patterns. More precisely, $L_1$, $L_8$, $L_9$, $L_{11}$, $L_{12}$, and $L_{14}$ move towards the center $(0,0)$, while all the other points move away from the center. Our analysis suggests that $L_1$, $L_8$, $L_9$, $L_{11}$, $L_{12}$, and $L_{14}$ collide with the origin when $\mu = 1$. We observe that all the equilibrium points evolve along the axes of symmetry $y = 0$, $y = \sqrt{3}$ and $y = - \sqrt{3}$. At this point, we would like to note that the centers of all the primary bodies are completely unaffected by the shift of the mass parameter.

Knowing the exact positions $(x_0,y_0)$ of the equilibrium points, we can easily determine their linear stability, through the nature of the four roots of the characteristic equation. Our computations indicate that the vast majority of the libration points are always unstable, when the mass parameter $\mu$ varies in the interval $(0,1]$. Only $L_3$, $L_4$, and $L_5$ can be stable however only for extremely small values of the mass parameter and especially when $\mu \in (0, 0.00226341]$.

\section{The basins of attraction}
\label{bas}

There is no doubt that the most well-known numerical method for solving systems of nonlinear equations is the famous Newton-Raphson method. This method is applicable to systems of multivariate functions $f({\bf{x}}) = 0$ through the iterative scheme
\begin{equation}
{\bf{x}}_{n+1} = {\bf{x}}_{n} - J^{-1}f({\bf{x}}_{n}),
\label{sch}
\end{equation}
where $f({\bf{x_n}})$ denotes the system of equations, while $J^{-1}$ is the corresponding inverse Jacobian matrix. In our case Eqs. (\ref{lps}) describe the system of the differential equations.

Decomposing the above-mentioned iterative scheme we obtain the following iterative formulae for each coordinate
\begin{align}
x_{n+1} &= x_n - \left( \frac{\Omega_x \Omega_{yy} - \Omega_y \Omega_{xy}}{\Omega_{yy} \Omega_{xx} - \Omega^2_{xy}} \right)_{(x_n,y_n)}, \nonumber\\
y_{n+1} &= y_n + \left( \frac{\Omega_x \Omega_{yx} - \Omega_y \Omega_{xx}}{\Omega_{yy} \Omega_{xx} - \Omega^2_{xy}} \right)_{(x_n,y_n)},
\label{nrm}
\end{align}
where $x_n$, $y_n$ are the values of the $x$ and $y$ coordinates at the $n$-th step of the iterative process.

The philosophy behind the Newton-Raphson method is the following: The numerical code is activated when an initial condition $(x_0,y_0)$, on the configuration plane, is given, while the iterative procedure continues until an equilibrium point (attractor) is reached, with the desired predefined accuracy. If the particular initial condition leads to one of the libration points of the system it means that the numerical method converges for that particular initial condition. At this point, it should be emphasized that in general terms the method does not converge equally well for all the available initial conditions. The sets of the initial conditions which lead to the same attractor compose the so-called Newton-Raphson basins of attraction or basins of convergence or even attracting domains/regions. Nevertheless, it should be clarified that the Newton-Raphson basins of convergence should not be mistaken, by no means, with the basins of attractions which are present in systems with dissipation.

From the iterative formulae of Eqs. (\ref{nrm}) it becomes evident that they should reflect some of the most intrinsic properties of the Hamiltonian system. This is true if we take into account that they contain the derivatives of both first and second order of the effective potential function $\Omega(x,y)$.

A double scan of the configuration $(x,y)$ plane is performed for revealing the structures of the basins of attraction. In particular, a dense uniform grid of $1024 \times 1024$ $(x_0,y_0)$ nodes is defined which shall be used as initial conditions of the iterative scheme. Evidently, the initial conditions of the centers of the primary bodies are of course excluded from all the grids because for these initial conditions the distances $r_i$, $i = 0,...,3$ to the respective primaries are equal to zero and therefore several terms, entering formulae (\ref{nrm}), become singular. The number $N$ of the iterations, required for obtaining the desired accuracy, is also monitored during the classification of the nodes. For our computations, the maximum allowed number of iterations is $N_{\rm max} = 500$, while the iterations stop only when an attractor is reached, with accuracy of $10^{-15}$.

In the following subsections we will determine how the mass parameter $\mu$ affects the structure of the Newton-raphson basins of convergence in the CR5BP, by considering two cases regarding the total number of the attractors (equilibrium points). For the classification of the nodes on the configuration $(x,y)$ plane we will use color-coded diagrams (CCDs), in which each pixel is assigned a different color, according to the final state (attractor) of the corresponding initial condition.

\subsection{Case I: Nine equilibrium points}
\label{ss1}

\begin{figure*}[!t]
\centering
\resizebox{\hsize}{!}{\includegraphics{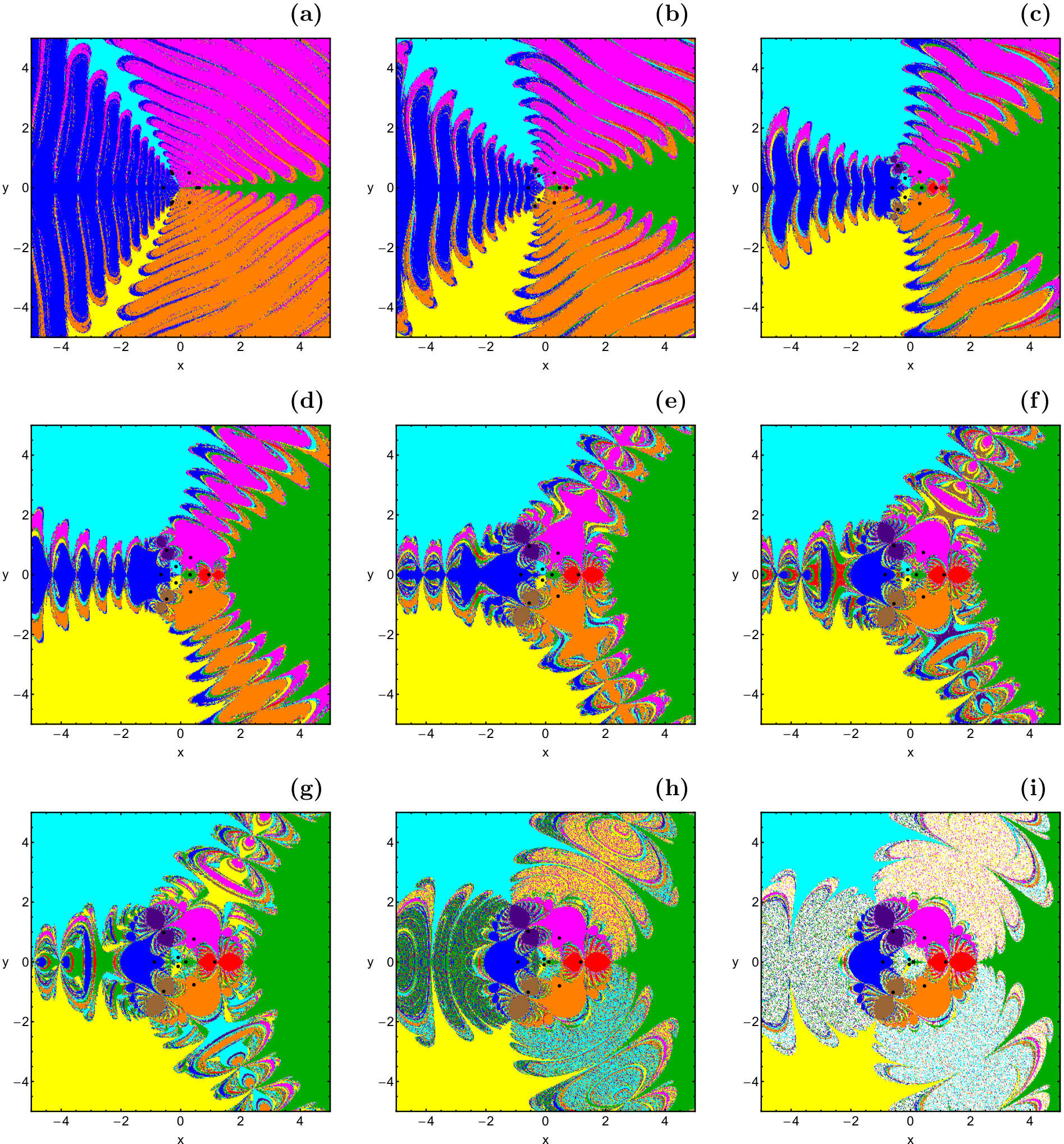}}
\caption{The Newton-Raphson basins of attraction on the configuration $(x,y)$ plane for the first case, where nine equilibrium points are present. (a): $\mu = 0.001$; (b): $\mu = 0.03$; (c): $\mu = 0.2$; (d): $\mu = 0.4$; (e): $\mu = 0.8$; (f): $\mu = 0.85$; (g): $\mu = 0.9$; (h): $\mu = 0.98$; (i): $\mu = 0.986172$. The positions of the equilibrium points are indicated by black dots. The color code, denoting the nine attractors, is as follows: $L_1$ (green); $L_2$ (red); $L_3$ (blue); $L_4$ (magenta); $L_5$ (orange); $L_6$ (indigo); $L_7$ (brown); $L_8$ (cyan); $L_9$ (yellow); non-converging points (white). (Color figure online).}
\label{c1}
\end{figure*}

\begin{figure*}[!t]
\centering
\resizebox{\hsize}{!}{\includegraphics{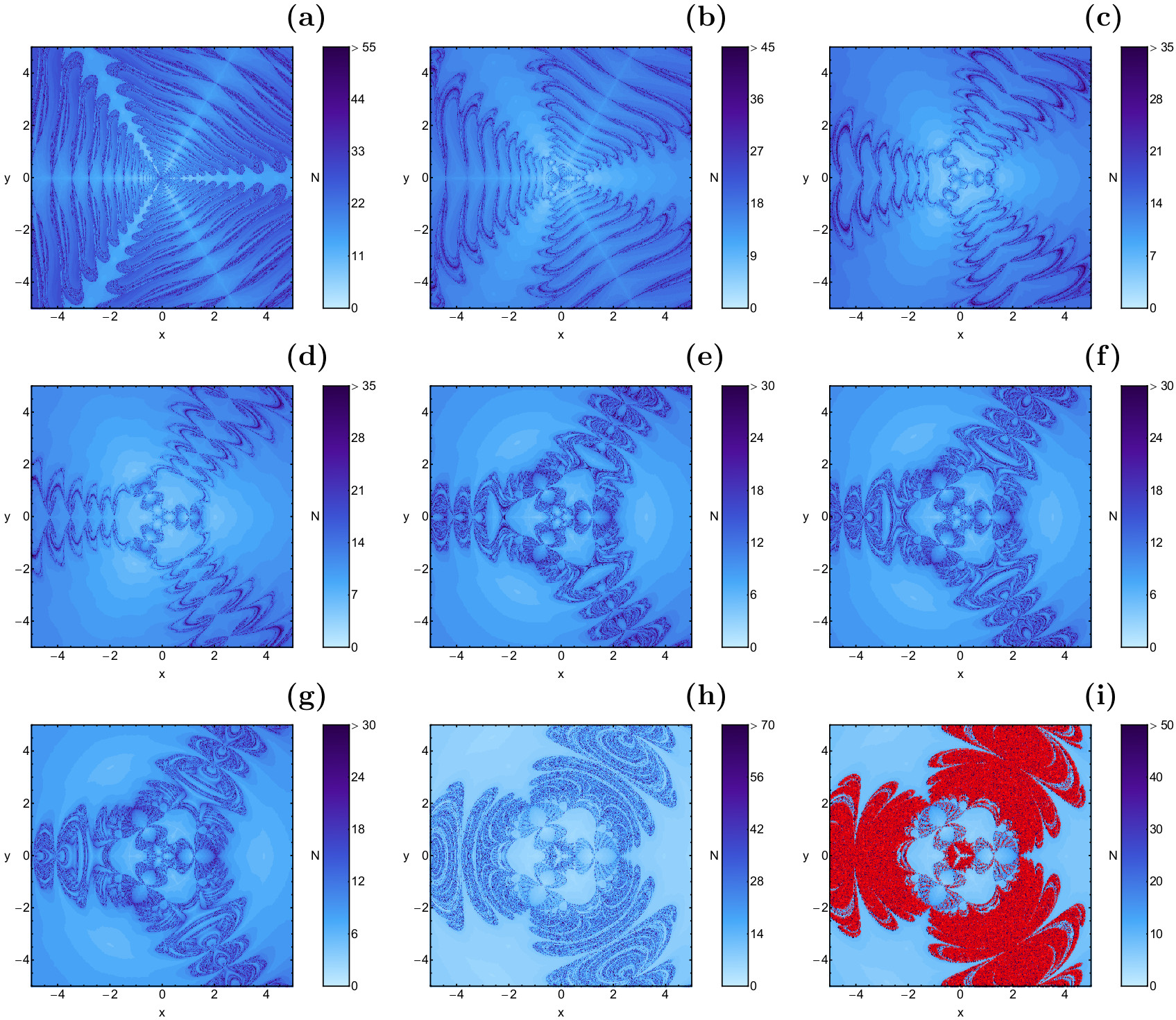}}
\caption{The distribution of the corresponding number $N$ of required iterations for obtaining the Newton-Raphson basins of attraction shown in Fig. \ref{c1}(a-i). The slow converging points are shown in red. (Color figure online).}
\label{n1}
\end{figure*}

\begin{figure*}[!t]
\centering
\resizebox{\hsize}{!}{\includegraphics{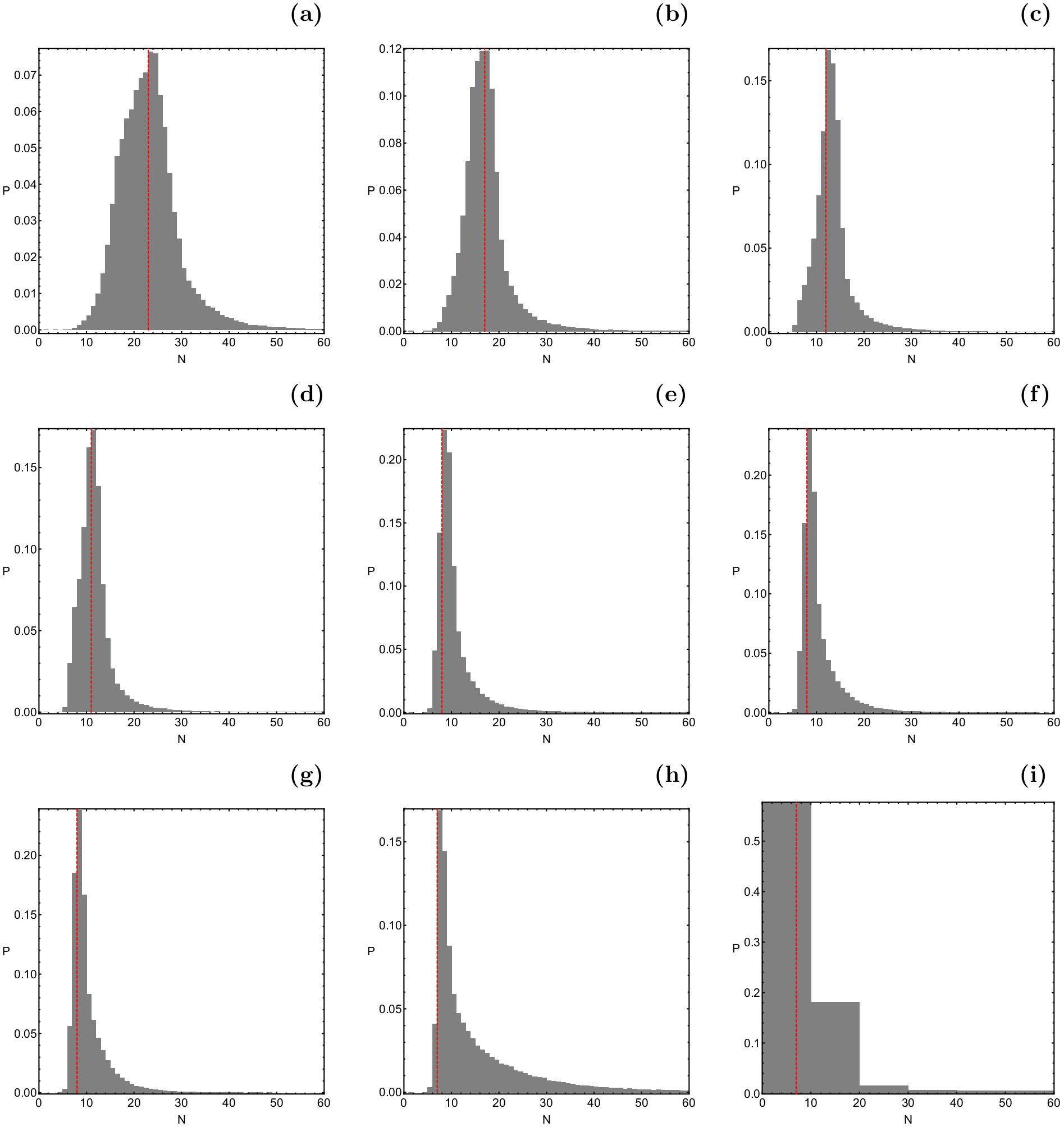}}
\caption{The corresponding probability distribution of required iterations for obtaining the Newton-Raphson basins of attraction shown in Fig. \ref{c1}(a-i). The vertical, dashed, red line indicates, in each case, the most probable number $N^{*}$ of iterations. (Color figure online).}
\label{p1}
\end{figure*}

\begin{figure*}[!t]
\centering
\resizebox{0.7\hsize}{!}{\includegraphics{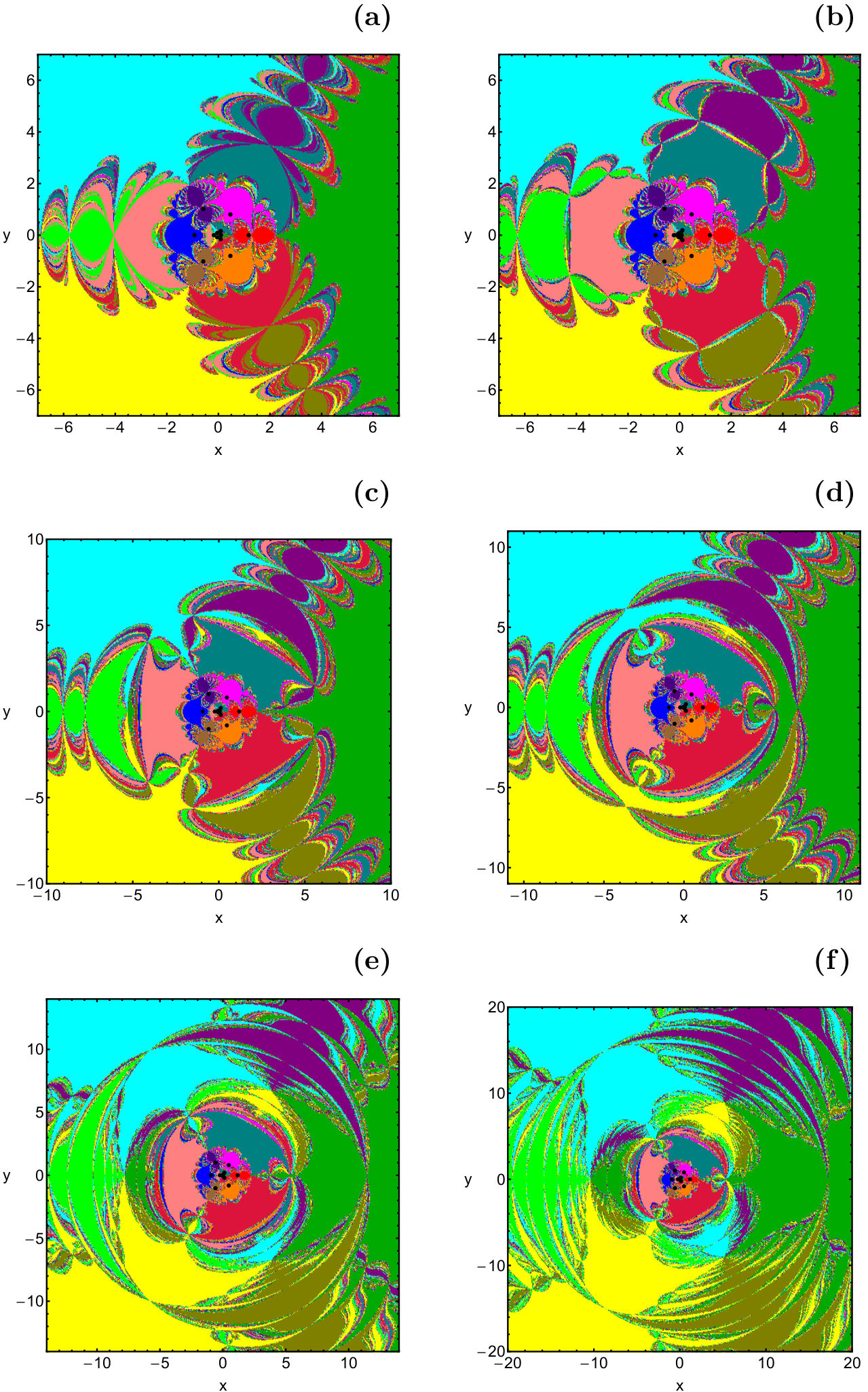}}
\caption{The Newton-Raphson basins of attraction on the configuration $(x,y)$ plane for the second case, where fifteen equilibrium points exist. (a): $\mu = 0.986173$; (b): $\mu = 0.992$; (c): $\mu = 0.998$; (d): $\mu = 0.999$; (e): $\mu = 0.9999$; (f): $\mu = 0.99999$. The positions of the equilibrium points are indicated by black dots. The color code, denoting the nine attractors, is as follows: $L_1$ (green); $L_2$ (red); $L_3$ (blue); $L_4$ (magenta); $L_5$ (orange); $L_6$ (indigo); $L_7$ (brown); $L_8$ (cyan); $L_9$ (yellow); $L_{10}$ (pink); $L_{11}$ (lime); $L_{12}$ (purple); $L_{13}$ (olive); $L_{14}$ (teal); $L_{15}$ (crimson); non-converging points (white). (Color figure online).}
\label{c2}
\end{figure*}

\begin{figure*}[!t]
\centering
\resizebox{0.9\hsize}{!}{\includegraphics{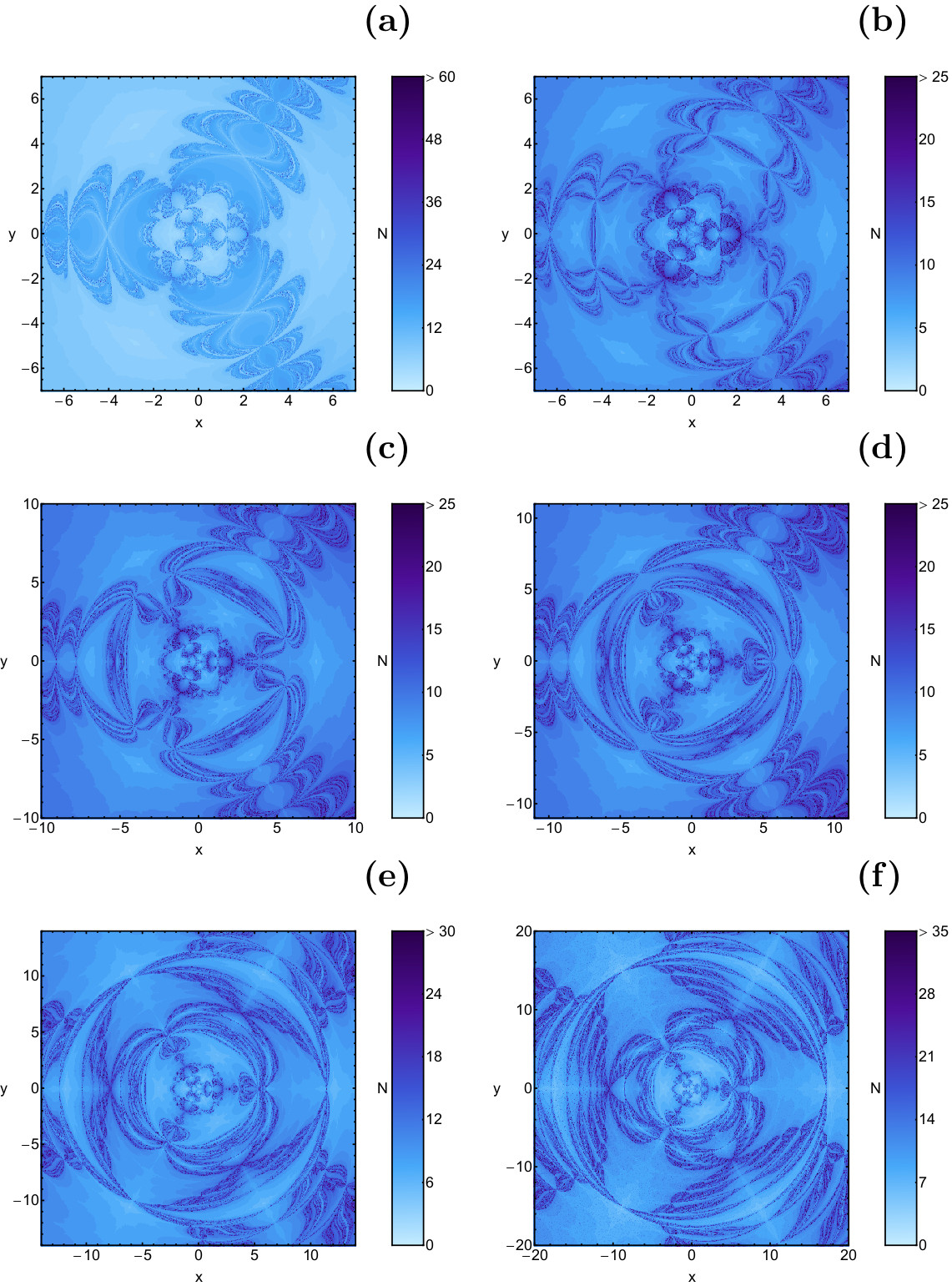}}
\caption{The distribution of the corresponding number $N$ of required iterations for obtaining the Newton-Raphson basins of attraction shown in Fig. \ref{c2}(a-f). (Color figure online).}
\label{n2}
\end{figure*}

\begin{figure*}[!t]
\centering
\resizebox{0.8\hsize}{!}{\includegraphics{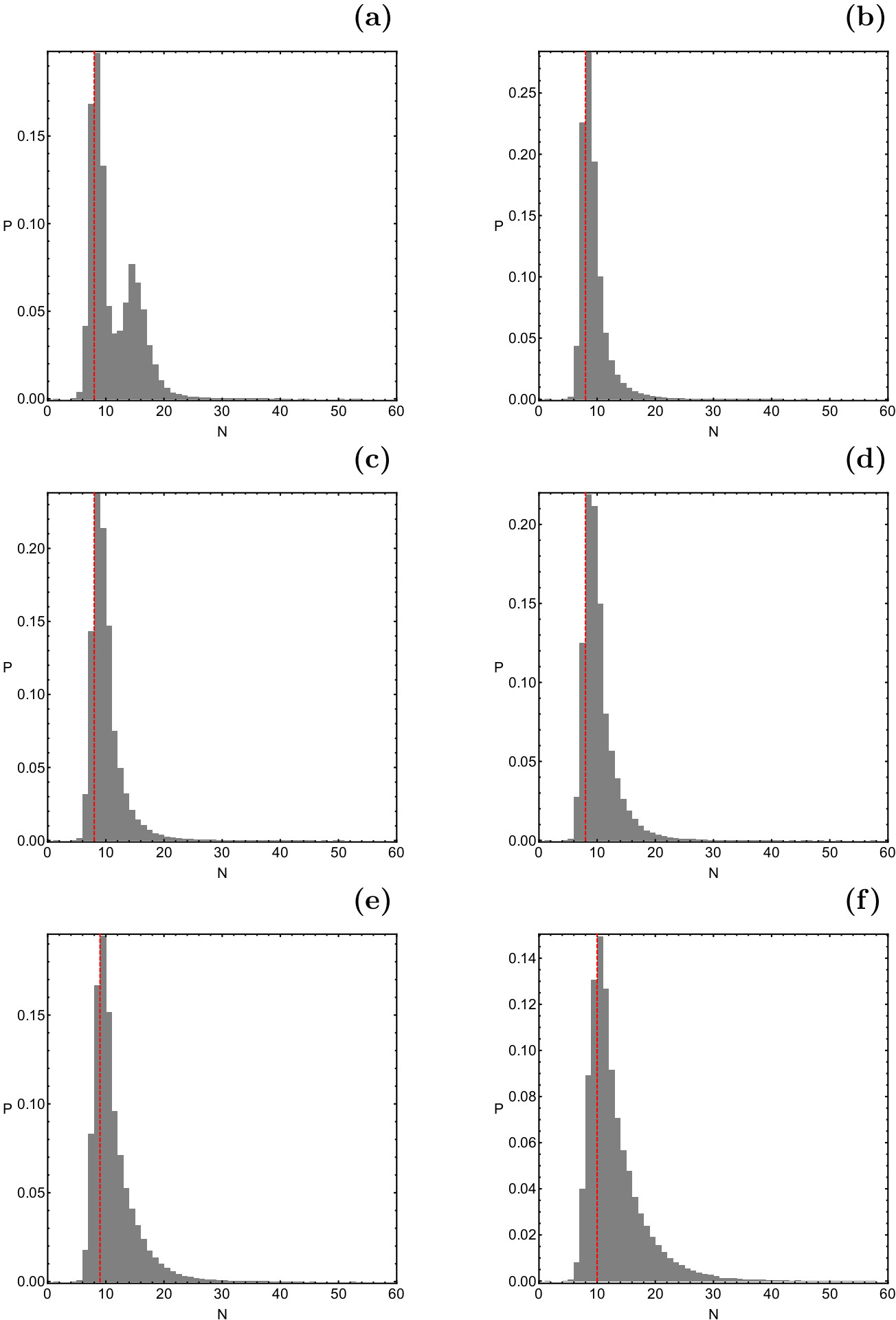}}
\caption{The corresponding probability distribution of required iterations for obtaining the Newton-Raphson basins of attraction shown in Fig. \ref{c2}(a-f). The vertical, dashed, red line indicates, in each case, the most probable number $N^{*}$ of iterations. (Color figure online).}
\label{p2}
\end{figure*}

\begin{figure*}[!t]
\centering
\resizebox{\hsize}{!}{\includegraphics{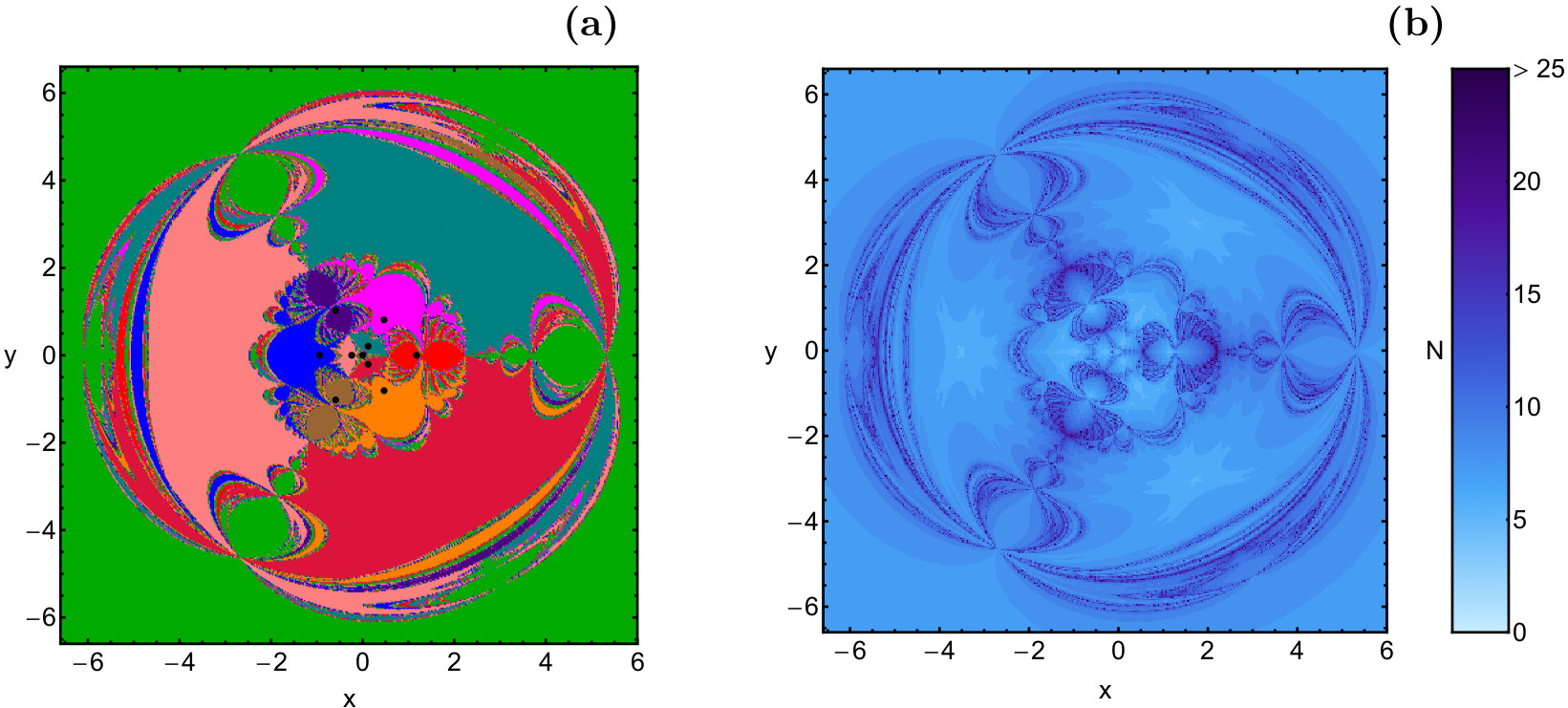}}
\caption{(a-left): The Newton-Raphson basins of attraction on the configuration $(x,y)$ plane when the central primary body is absent $(\mu = 1)$. The color code is the same as in Fig. \ref{c2}. (b-right): The distribution of the corresponding number $N$ of required iterations. (Color figure online).}
\label{sm}
\end{figure*}

\begin{figure*}[!t]
\centering
\resizebox{\hsize}{!}{\includegraphics{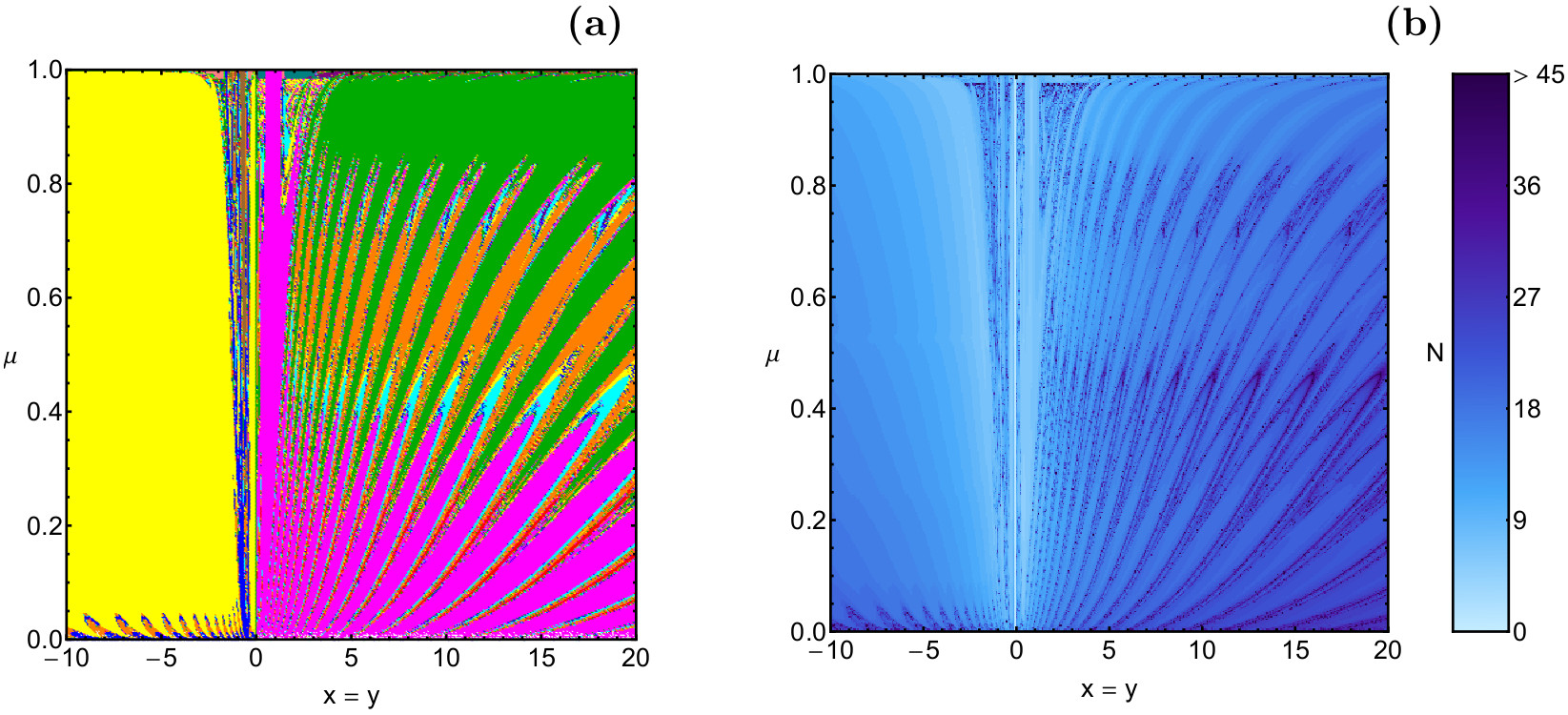}}
\caption{(a-left): The Newton-Raphson basins of attraction on the $(x = y,\mu)$ plane, when $\mu \in (0,1]$. The color code denoting the attractors is the same as in Figs. \ref{c1} and \ref{c2}. (b-right): The distribution of the corresponding number $N$ of required iterations for obtaining the basins of convergence shown in panel (a). (Color figure online).}
\label{xym}
\end{figure*}

\begin{figure*}[!t]
\centering
\resizebox{\hsize}{!}{\includegraphics{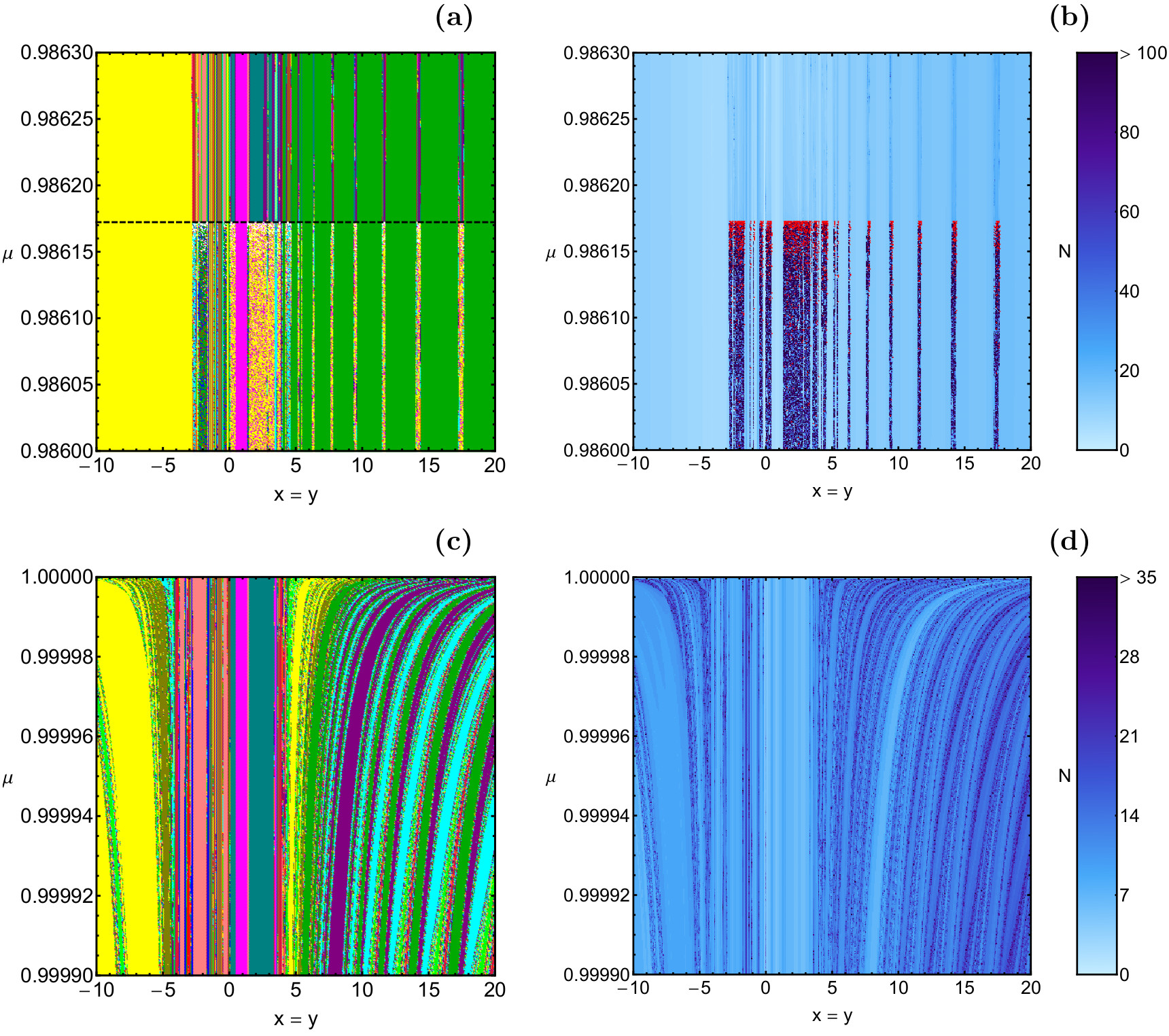}}
\caption{Magnification of the $(x = y,\mu)$ plane, around (a-upper left): the critical value of the mass parameter $\mu^{*}$ (horizontal, black, dashed line) and (c-lower left): $\mu = 1$. The color code denoting the attractors is the same as in Figs. \ref{c1} and \ref{c2}. The corresponding distributions of the required iterations are shown in panel (b) and (d), respectively. (Color figure online).}
\label{xym2}
\end{figure*}

Our first case under investigation considers the scenario where nine equilibrium points exist, that is when $0 < \mu \leq 0.98617275$. The evolution of the geometry of the basins of attraction, for nine values of the mass parameter, is illustrated in Fig. \ref{c1}(a-i). It is seen that in all cases the configuration $(x,y)$ plane is covered by several well-defined basins of convergence which extend to infinity. Furthermore, in the vicinity of the basin boundaries we observe a highly fractal\footnote{By the term fractal we simply mean that the particular area has a fractal-like geometry, without conducting any additional calculations, until for now, as in \citet{AVS01}.} mixture of initial conditions. This means that the basin boundaries are highly chaotic and therefore the final state (attractor) of an initial condition inside this area is highly sensitive. In particular, even the slightest change of the initial conditions automatically leads to a completely different attractor. Therefore for the initial conditions in the basin boundaries it is almost impossible to predict their final state (equilibrium point).

With increasing value of the mass parameter $\mu$ the structure of the configuration $(x,y)$ plane changes drastically. The most noticeable changes can be summarized as follows:
\begin{itemize}
  \item The extent of the attracting domains, corresponding to equilibrium points $L_3$, $L_4$ and $L_5$, decreases.
  \item The area of the basins of convergence, corresponding to libration points $L_2$, $L_6$ and $L_7$, increases.
  \item The extent of the basins of attraction, corresponding to attractors $L_1$, $L_8$ and $L_9$, initially increases, while for $\mu > 0.9$ the tendency is reversed.
\end{itemize}

In panel (i) of Fig. \ref{c1}, where $\mu = 0.986172$, we identify a considerable amount of non-converging initial conditions, which mainly surround the central region. However, additional numerical calculations revealed that these initial conditions are in fact extremely slow converging points, which do converge to one of the nine attractors, only after a considerable amount of iterations $(N \gg 500)$.

In Fig. \ref{n1}(a-i) we present the corresponding number $N$ of iterations, using tones of blue. It is seen that initial conditions inside the basins of attraction converge relatively fast $(N < 20)$, while the slowest converging points $(N > 20)$ are those in the vicinity of the basin boundaries. The corresponding probability distribution of the required iterations is given in Fig. \ref{p1}(a-i). In all plots the tails of the histograms extend so as to cover 95\% of the corresponding distribution of iterations. The definition of the probability $P$ is the following: if $N_0$ initial conditions $(x_0,y_0)$ converge, after $N$ iterations, to one of the libration points then $P = N_0/N_t$, where $N_t$ is the total number of nodes in every CCD. Our analysis suggest that the most probable number $N^{*}$ of iterations (see the red vertical dashed lines in Fig. \ref{p1}) constantly reduces from $N^{*} = 23$, when $\mu = 0.001$ to $N^{*} = 7$, when $\mu = 0.986172$.

\subsection{Case II: Fifteen equilibrium points}
\label{ss2}

In this case, where $\mu^{*} \leq \mu < 1$, there are fifteen equilibrium points: five on the $x$ axis and ten with $y \neq 0$. The Newton-Raphson basins of attraction for six values of the mass parameter are given in Fig. \ref{c2}(a-f). It is seen that the basins of attraction, corresponding to libration points $L_i$, $i = 10,...,15$, have the shape of butterfly wings and they also extend to infinity, as those of the equilibrium points $L_i$, $i = 1,...,9$, discussed in the previous subsection. Furthermore, we observe that the entire pattern (the overall structure on the configuration plane composed of all the different attracting domains) grows rapidly as the mass parameter tends to 1.

It was found that just above the critical value $\mu^{*}$, that is when $\mu = 0.986173$ (see panel (a) in Fig. \ref{c2}), a portion of initial conditions fails to obtain the desired accuracy. In particular, for these initial conditions the iterative scheme reaches an attractor (equilibrium point) with an accuracy of $10^{-14}$ and then the convergence stops, even if we allow the iterative procedure to continue for more than 500 iterations. We believe that this numerical malfunction of the code is strongly related with the fact that we are just above the critical value of the mass parameter, where the basin boundaries are very fractalized. Nevertheless, we count all these initial conditions as regular converging nodes, taking into account that an accuracy of $10^{-14}$ is also sufficient and therefore acceptable.

The distribution of the corresponding number $N$ of iterations, required for obtaining the desired accuracy, is illustrated in Fig. \ref{n2}(a-f). We see that in most cases the vast majority of the initial conditions converge to one of the attractors (equilibrium points) within the first 25 iterations. The corresponding probability distributions are given in Fig. \ref{p2}(a-f). In this case the most probable number $N^{*}$ of iterations increases with increasing value of the mass parameter. Indeed, for $\mu = 0.986173$ we have that $N^{*} = 8$, while for $\mu = 0.99999$ the value of $N^{*}$ is elevated to 10.

When $\mu = 1$ (which means that the central primary body is missing) the five body problem degenerates to the equilateral restricted four-body problem. This fact is also verified through the corresponding Newton-Raphson basins of convergence. In panel (a) of Fig. \ref{sm} we provide the CCD for $\mu = 1$. One can easily observe that the overall structure is completely identical to that of Fig. 5 of \citet{Z17a}, where the attracting domains of the planar equilateral restricted four-body problem had been investigated.

\subsection{An overview analysis}
\label{geno}

\begin{figure}[!t]
\centering
\resizebox{\hsize}{!}{\includegraphics{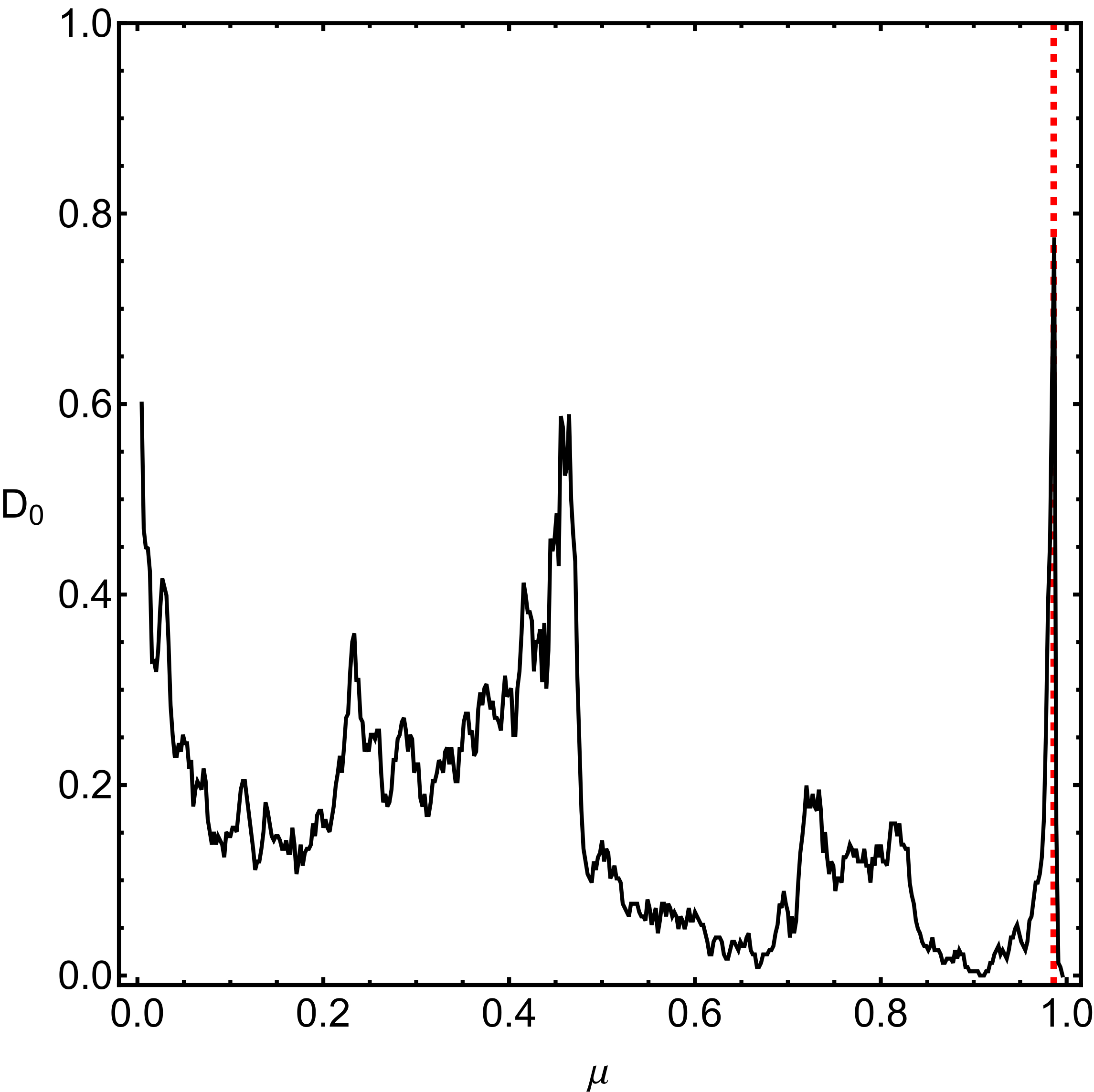}}
\caption{Evolution of the fractal dimension $D_0$ of the $(x = y,\mu)$ plane of panel (a) of Fig. \ref{xym} as a function of the mass parameter $\mu$. $D_0 = 1$ implies total fractality, while $D_0 = 0$ means zero fractality. The red, dashed, vertical line indicates the critical value of the mass parameter $(\mu^{*})$, which distinguishes between the two cases, regarding the total number and the type of the equilibrium points. (Color figure online).}
\label{fract}
\end{figure}

The color-coded convergence diagrams on the configuration $(x,y)$ space, presented earlier in Figs. \ref{c1} and \ref{c2} provide sufficient information regarding the attracting domains, however for only a fixed value of the mass parameter $\mu$. In order to overcome this handicap we can define a new type of distribution of initial conditions which will allow us to scan a continuous spectrum of $\mu$ values, rather than few discrete levels. The most interesting configuration is to set $x = y$, while the value of the mass parameter will vary in the interval $(0,1]$. This technique allows us to construct, once more, a two-dimensional plane in which the $x$ or the $y$ coordinate is the abscissa, while the value of $\mu$ is always the ordinate. Panel (a) of Fig. \ref{xym} shows the basins of attraction on the $(x = y,\mu)$ plane, while in panel (b) of the same figure the distribution of the corresponding number $N$ of required iterations for obtaining the Newton-Raphson basins of convergence is shown. In panel (a) of Fig. \ref{xym} it can be seen very clearly how the convergence properties of the system change when $\mu \geq \mu^{*}$.

It would be very interesting to know what happens around the critical value of the mass parameter $\mu^{*}$, where the number of the convergence basins changes from 9 to 15. In the same vein, useful results could be obtained near $\mu = 1$, where the dynamical properties of the system change drastically, as it degenerates to the restricted four-body problem. In panels (a) and (c) of Fig. \ref{xym2} we present magnifications of panel (a) of Fig. \ref{xym}, near the two regions with high interest. Now the transition is much more clear, while the changes on the convergence properties of the $(x = y,\mu)$ plane are more evident. The corresponding distributions of the required iterations are given in panels (b) and (d) of Fig. \ref{xym2}, respectively.

So far in our work, the degree of fractality of the several types of two-dimensional convergence planes has been discussed only in a qualitative way. We seen that the highly fractal domains are those mainly located in the vicinity of the basin boundaries in which it is almost impossible to predict the final states of the initial conditions. Inside the convergence areas on the other hand, we can easily predict from which attractor (equilibrium point) each initial condition is attracted by. On this basis, it would be an issue of great importance to provide quantitative results about the degree of fractality of the $(x = y,\mu)$ plane of panel (a) of Fig. \ref{xym}. For measuring the degree of fractality we calculated the uncertainty dimension \citep{O93} for different values of the mass parameter $\mu$, thus following the computational approach discussed in \citet{AVS01}. At this point, it should be emphasized that the uncertainty dimension is completely independent of the particular set of initial conditions used for its calculation.

In Fig. \ref{fract} we depict the evolution of the uncertainty dimension $D_0$ of the $(x = y,\mu)$ plane, as a function of the mass parameter $\mu$. It is seen that $D_0 \in (0,1)$ because the calculation of $D_0$ was performed for only a ``1D slice'' of initial conditions. It is interesting to note that as the numerical value of $\mu$ varies in the interval $(0,1]$ the degree of fractality fluctuates, while the highest value of $D_0$ is observed exactly at the critical value of the mass parameter. On the contrary, $D_0$ is almost zero (display the lowest possible value) when $\mu = 1$, that is when the system degenerates to the circular restricted four-body problem.

\section{Parametric evolution of the basin entropy}
\label{bee}

In \citet{DWGGS16} a new tool for measuring the uncertainty of the basins has been introduced. This new tool is called the ``basin entropy" and refers to the topology of the basins, thus describing the notion of fractality and unpredictability in the context of basins of attraction or basins of escape.

\begin{figure*}[!t]
\centering
\resizebox{\hsize}{!}{\includegraphics{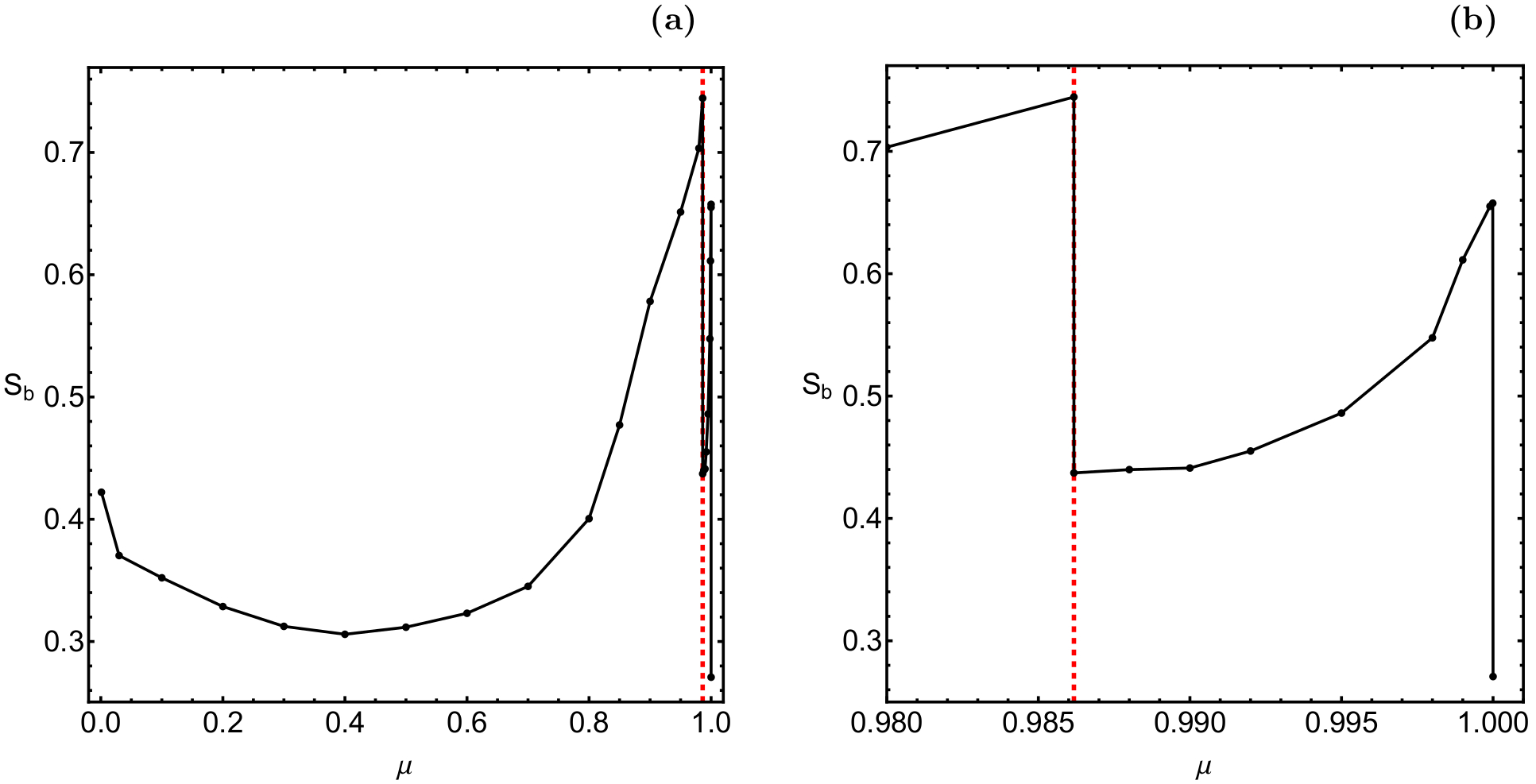}}
\caption{(a-left): Evolution of the basin entropy $S_b$, of the configuration $(x,y)$ plane, as a function of the mass parameter $\mu$. The red, dashed, vertical line indicates the critical value $\mu^{*}$. (b-right): Magnification of panel (a) near the critical value of the mass parameter. (Color figure online).}
\label{be}
\end{figure*}

Let us briefly recall the numerical algorithm of the basin entropy. We assume that there are $N(A)$ attractors (equilibrium points) in a certain region $R$ of the configuration plane in our dynamical systems. Moreover, $R$ can be subdivided into a grid composed of $N$ square boxes. Each box of the square grid can contain between 1 and $N(A)$ attractors. Therefore we can denote $P_{i,j}$ the probability that inside the box $i$ the resulting attractor is $j$. Due to the fact that inside the box the initial conditions are completely independent, the Gibbs entropy, of every box $i$, is given by
\begin{equation}
S_{i} = \sum_{j=1}^{m_{i}}P_{i,j}\log_{10}\left(\frac{1}{P_{i,j}}\right),
\end{equation}
where $m_{i} \in [1,N_{A}]$ is the number of the attractors inside the box $i$.

The entropy of the entire region $R$, on the configuration plane, can be computed as the sum of the entropies of the resulting $N$ boxes of the square grid as $S = \sum_{i=1}^{N} S_{i}$. On this basis, the entropy relative to the total number of boxes $N$, which is called basin entropy $S_{b}$, is given explicitly by the following expression
\begin{equation}
S_{b} = \frac{1}{N}\sum_{i=1}^{N}\sum_{j=1}^{m_{i}}P_{i,j}\log_{10}\left(\frac{1}{P_{i,j}}\right).
\end{equation}

Using the above-mentioned expressions and also adopting the value $\varepsilon = 0.005$, suggested in \citet{DWGGS16}, we computed the basin entropy $S_{b}$ of the configuration plane, when the mass parameter lies in the interval $\mu \in (0,1]$. Here it should be clarified that in the case where non-converging points are present we count them as an additional type of basin which coexists with the other basins, corresponding to the equilibrium points. In Fig. \ref{be} we present the evolution of the basin entropy as a function of the mass parameter. At this point, it should be noted that for creating this diagram we used numerical results not only for the cases presented earlier in Figs. \ref{c1} and \ref{c2} but also from additional values of $\mu$.

We see that as long as $\mu > 0$ the basin entropy decreases up to about $\mu = 0.4$, while for higher values of the mass parameter the tendency is reversed. The highest value of $S_b$ is observed at the critical value $\mu^{*}$, while the basin entropy displays a sudden drop when $\mu = 0.986173$. However with increasing value of $\mu$ it continues to gradually grows as the mass parameter tends to one. Interestingly, at exactly $\mu = 1$ (which corresponds to an entire different dynamical system) the basin entropy is suddenly reduced to minimum.

It would be very useful if we could directly relate the results of Fig. \ref{fract}, regarding the fractal dimension $D_0$ with those of Fig. \ref{be}, about the basin entropy. Unfortunately, the results of the two figures correspond to entirely different types of distributions of initial conditions. In particular, for computing the fractal dimension we used information from the $(x = y,\mu)$ plane (that is a one-dimensional set of initial conditions), while for the calculation of the basin entropy we exploited information from the two-dimensional sets of initial conditions on the configuration $(x,y)$ plane. Nevertheless, we cannot ignore the fact that both dynamical quantities (fractal dimension and basin entropy) suggest that the highest degree of fractalization is observed exactly at the critical value of the mass parameter. This implies that for this particular value of $\mu$ there is almost complete lack of predictability in a totally deterministic Hamiltonian system.

\section{Discussion and conclusions}
\label{conc}

We numerically explored the basins of convergence, related to the equilibrium points, in the planar circular restricted five-body problem. More precisely, we demonstrated how the mass parameter $\mu$ influences the position as well as the linear stability of the libration points. The multivariate version of the Newton-raphson iterative scheme was used for revealing the corresponding basins of attraction on the configuration $(x,y)$ plane. These attracting domains play a significant role, since they explain how each point of the configuration plane is attracted by the libration points of the system, which act, in a way, as attractors. We managed to monitor how the Newton-Raphson basins of attraction evolve as a function of the mass parameter. Another important aspect of this work was the relation between the basins of convergence and the corresponding number of required iterations and the respective probability distributions.

To our knowledge this is the first time that the Newton-Raphson basins of attraction in the planar circular restricted five-body problem are numerically investigated in a systematic manner. On this basis, the presented results are novel and this is exactly the contribution of our work.

The following list contains the most important conclusions of our numerical analysis.
\begin{enumerate}
  \item The stability analysis suggests that all most of the equilibrium points of the system are always linearly unstable. Only $L_3$, $L_4$, and $L_5$ can be linearly stable and only for extremely small values of the mass parameter.
  \item The attracting domains, associated to all the libration points, extend to infinity, in all studied cases. Moreover, the $2\pi/3$ symmetry of the system is present in all the convergence diagrams on the $(x,y)$ plane.
  \item Just before the critical value $\mu^{*}$ we detected the existence of non-converging initial conditions. Additional computations revealed that these initial conditions are in fact extremely slow converging points, which do converges to one of the equilibrium points of the system but only after many iterations $(N \gg 500)$.
  \item Through the classification of the nodes on the several two-dimensional planes we did not encounter any true non-converging initial conditions.
  \item The multivariate Newton-Raphson method was found to converge very fast $(0 \leq N < 10)$ for initial conditions close to the equilibrium point, fast $(10 \leq N < 20)$ and slow $(20 \leq N < 40)$ for initial conditions that complement the central regions of the very fast convergence, and very slow $(N \geq 40)$ for initial conditions of dispersed points lying either in the vicinity of the basin boundaries, or between the dense regions of the libration points.
  \item As the value of the mass parameter increases from 0 to $\mu^{*}$ the most probable number of required iterations, $N^{*}$, was found to decrease, while for $\mu > \mu^{*}$ the tendency is reversed.
  \item Both the fractal dimension and the basin entropy were found to display their global maximum at exactly the critical value of the mass parameter $\mu^{*}$. Therefore we have a strong numerical evidence that for this value the system displays its highest degree of fractalization. On the other hand, for $\mu = 1$ we detected the lowest possible value for both of them.
\end{enumerate}

A double precision numerical code, written in standard \verb!FORTRAN 77! \citep{PTVF92}, was used for the classification of the initial conditions into the different basins of attraction. In addition, for all the graphical illustration of the paper we used the latest version 11.2 of Mathematica$^{\circledR}$ \citep{W03}. Using an Intel$^{\circledR}$ Quad-Core\textsuperscript{TM} i7 2.4 GHz PC the required CPU time, for the classification of each set of initial conditions, was about 5 minutes.

\section{Future work}
\label{fut}

In a series of previous papers we numerically investigated the Newton-Raphson basins of convergence in the circular restricted three-body problem \citep{Z16a}, as well as in the restricted four-body problem \citep{SAA17,Z17a}. It is in our future plans to devote a new paper in order to demonstrate in detail all the similarities and differences between the several cases, in an attempt to obtain a general overview regarding the convergence properties and the fractal structures, associated to the circular restricted $N$-body problem, with $N = 3,4,5$.

In systems where three or more basins of convergence (or escape) coexist one should know whether these basins verify, or not, the so-called Wada property, that is where three or more basins share the same boundary \citep[e.g.,]{PCOG96}. An easy way of determining if the Wada property is verified is by plotting the unstable manifold, of a periodic orbit found in the boundary, and monitoring which basins are crossed by it. These calculation, which will allow us to obtain additional relevant information regarding the predictability of the system, will be also conduced in the future paper.

Another interesting aspect would be to use other types of iterative schemes and compare the corresponding similarities and differences related to the basins of attraction of the equilibrium points of the dynamical system. More precisely, we could use iterative methods of higher order, with respect to the classical Newton-Raphson method of second order. In a recent paper \citep{Z17b} we deployed a large collection of iterative schemes, of higher order, for revealing and therefore comparing the convergence properties of the Hill problem with oblateness and radiation pressure. So far, all these iterative schemes work only for solving an equation with one variable. Currently, we are trying to expand all these numerical methods so as to be able to use them for solving a system of two nonlinear equations (such as that of Eq. (\ref{lps})).

We hope that all the above-mentioned ideas would lead to useful, and perhaps unexpected, results in the very active field of attracting domains of equilibrium points in dynamical systems.

\section*{Acknowledgments}
\footnotesize

The authors would like to express their warmest thanks to the anonymous referees for the careful reading of the manuscript and for all the apt suggestions and comments which allowed us to improve both the quality and the clarity of the paper.

\section*{Compliance with Ethical Standards}

\begin{itemize}
  \item Funding: The authors states that they have not received any research grants.
  \item Conflict of interest: The authors declare that they have no conflict of interest.
\end{itemize}

\end{document}